# Generalized Richardson-Lucy (GRL) for analyzing multi-shell diffusion MRI data

Fenghua Guo[1], Alexander Leemans[1], Max A. Viergever[1], Flavio Dell'Acqua[2], Alberto De Luca[1]

[1] Image Sciences Institute, University Medical Center Utrecht, Utrecht University, the Netherlands

[2] NatBrainLab, Department of Forensics and Neurodevelopmental Sciences, Sackler Institute for Translational Neurodevelopment, King's College London, UK

## Abstract

Spherical deconvolution is a widely used approach to quantify the fiber orientation distribution (FOD) from diffusion MRI data of the brain. The damped Richardson-Lucy (dRL) is an algorithm developed to perform robust spherical deconvolution on single-shell diffusion MRI data while suppressing spurious FOD peaks due to noise or partial volume effects. Due to recent progress in acquisition hardware and scanning protocols, it is becoming increasingly common to acquire multi-shell diffusion MRI data, which allows for the modelling of multiple tissue types beyond white matter. While the dRL algorithm could, in theory, be directly applied to multi-shell data, it is not optimised to exploit its information content to model the signal from multiple tissue types. In this work, we introduce a new framework based on dRL – dubbed generalized Richardson-Lucy (GRL) – that uses multi-shell data in combination with user-chosen tissue models to disentangle partial volume effects and increase the accuracy in FOD estimation. Further, GRL estimates signal fraction maps associated to each user-selected model, which can be used during fiber tractography to dissect and terminate the tracking without need for additional structural data. The optimal weighting of multi-shell data in the fit and the robustness to noise and to partial volume effects of GRL was studied with synthetic data. Subsequently, we investigated the performance of GRL in comparison to dRL and to multi-shell constrained spherical deconvolution (MSCSD) on a high-resolution diffusion MRI dataset from the Human Connectome Project and on an MRI dataset acquired at 3T on a clinical scanner. In line with previous studies, we described the signal of the cerebrospinal-fluid and of the grey matter with isotropic diffusion models, whereas four diffusion models were considered to describe the white matter. With a third dataset including small diffusion weightings, we studied the feasibility of including intra-voxel incoherent motion effects due to blood pseudo-diffusion in the modelling. Further, the reliability of GRL was demonstrated with a test-retest scan of a subject acquired at 3T. Results of simulations show that GRL can robustly disentangle different tissue types at SNR above 20 with respect to the non-weighted image, and that it improves the angular accuracy of the FOD estimation as compared to dRL. On real data, GRL provides signal fraction maps that are physiologically plausible and consistent with those obtained with MSCSD, with correlation coefficients between the two methods up to 0.96. When considering IVIM effects, a high blood pseudo-diffusion fraction is observed in the medial temporal lobe and in the sagittal sinus. In comparison to dRL and MSCSD, GRL provided sharper FODs and less spurious peaks in presence of partial volume effects, but the FOD reconstructions are also highly dependent on the chosen modelling of white matter. When performing fiber tractography, GRL allows to terminate fiber tractography using the signal fraction maps, which results in a better tract termination at the grey-white matter interface or at the outer cortical surface. In terms of inter-scan reliability, GRL performed similarly to or better than compared methods. In conclusion, GRL offers a new modular and flexible framework to perform spherical deconvolution of multi-shell data.

## Key Words

# 1. Introduction

Diffusion MRI (dMRI) is an established technique to study the microstructure of brain white matter (WM). With fiber tractography (Mori et al., 1999), it is possible to reconstruct a 3D representation of the main WM bundles given knowledge on the direction of the underlying diffusion process. Fiber tractography uses indirect information from the diffusion process to attempt the reconstruction of the main WM bundles and has been shown to achieve good sensitivity, but also to produce large number of false positive reconstructions (Maier-Hein et al., 2019). When the process has a single diffusion orientation, the diffusion tensor imaging (DTI) method (Basser et al., 1994; Pierpaoli & Basser, 1996) can be employed to efficiently derive the main diffusion direction from six or more diffusion weighted measurements. However, the in-vivo brain is characterized by the presence of multiple fiber crossings (Behrens et al., 2007; Jeurissen et al., 2013, Dell'Acqua et al. 2013), which cannot be disentangled with DTI. To reconstruct the directions of multiple crossing fibers, more general approaches such as spherical deconvolution (Anderson, 2005; Dell'Acqua et al., 2010; Dell'Acqua and Tournier, 2019; Tournier et al., 2004), methods based on the kurtosis tensor (Glenn et al., 2015) or q-space inversion methods (Tuch et al., 2002; Wedeen et al., 2005) are commonly employed.

Spherical deconvolution approaches estimate the fiber orientation distribution (FOD) of WM using data acquired along multiple diffusion gradient orientations distributed over a sphere. When the acquisition is performed at a single diffusion weighting, the data is commonly referred as single-shell data, whereas multi-shell data implies that the sampling has been performed at two or more diffusion weightings. Once the FOD has been derived, it can be used directly to perform fiber tractography as well as for generalized voxel-wise fiber specific statistics (Dell'Acqua et al., 2013; D Raffelt et al., 2010).

Fiber tractography is the most common endpoint of FOD estimation methods. In the latest decade, fiber tractography applications are becoming a hot application in both research and clinical applications (Behrens et al., 2007; Daducci et al., 2014; Jones, 2010; Maier-Hein et al., 2019). One of the current flows of fiber tractography is that it is prone to produce false positive reconstructions (e.g. spurious fibers, (Maier-Hein et al., 2019)). To mitigate this issue, the fiber tractography community has invested large effort in devising advanced methods that discard false positive reconstructions, including particle filtering (Pontabry & Rousseau, 2011), probabilistic tractography with structural filtering (R. E. Smith et al., 2012), and global tractography (Christiaens et al., 2015). A complementary approach to improve the quality of fiber tractography is to improve the FOD estimation step, which is the aim of this work.

In general, a suggested data acquisition scheme for single shell spherical deconvolution approaches consists of at least 60 gradient directions with a diffusion weighting of around b-value = 3000 s/mm$^2$ (Tournier et al., 2013). Data at such strong diffusion weighting are generally characterized by degraded SNR (Descoteaux et al., 2011; Jones et al., 2013), but allows to resolve crossing fibers with higher angular resolution as compared to lower diffusion weightings. Several spherical deconvolution approaches (Dell'Acqua et al., 2007; Tournier et al., 2007) have been proposed to estimate the FOD from diffusion weighted signals. Among others, the Richardson-Lucy spherical deconvolution method (Dell'Acqua et al., 2007) is a well-established method to estimate the FOD from single-shell dMRI data. In a later formulation (Dell'Acqua et al., 2010), the approach was extended as damped Richardson-Lucy (dRL) to increase the robustness of the FOD estimation to partial volume effects with isotropic tissues and to prevent overfitting to noise.

Recent improvements in hardware have strongly increased the acquisition efficiency, making it feasible to collect dMRI data at multiple diffusion weighting (multi-shell) in a single acquisition within clinical achievable time. Multi-shell dMRI has potential to further improve the performance of the FOD estimation, as previously shown by Jeurissen et al. (Jeurissen et al., 2014) in the context of constrained spherical deconvolution (CSD) (Tournier et al., 2007), or by q-space sampling methods such as the Generalized q-Sampling Imaging (GQI) (Baete et al., 2016) and the Radial q-Space Sampling for diffusion spectrum imaging (RDSI) (Yeh et al., 2010).

In the multi-shell CSD (MS-CSD) study (Jeurissen et al., 2014), the authors modelled the three main tissue types of the brain (WM, grey matter (GM) and cerebrospinal-fluid (CSF)) to improve the accuracy of WM FODs in presence of partial volume effects in comparison to single shell CSD (Tournier et al., 2007), while also computing the signal fractions of each compartment. Different tissue types, as for instance WM, CSF and GM are characterized by distinct attenuation profiles over increasing diffusion weighting, which allows to model their contributions, a problem otherwise ill-posed with single shell data. Once signal fraction maps of different tissue classes have been derived, they could potentially be used to guide the fiber tractography procedure, in analogy to what is done with T1 derived segmentations (R. E. Smith et al., 2012). This unexplored application would not only avoid the extra-acquisition time of structural imaging, but also prevent residual misalignments between T1w data and dMRI. Further, the reconstruction of WM FODs could potentially benefit of multi-shell data by leveraging the high angular resolution of strong diffusion weighting in conjunction with the higher SNR of lower diffusion weighting signals.

In this study, we introduce the Generalized Richardson-Lucy (GRL) framework, a modular spherical deconvolution that aims at improving the reconstruction accuracy of WM FODs by accounting for an arbitrary – user defined – number of signal models representing different tissue types. We show through simulations that GRL can disentangle up to four different tissue types and accommodate multiple choices to represent the WM signal. Our analysis includes the use of two signal representations, such as DTI with a single set of eigenvalues, diffusion kurtosis imaging (DKI), and a microstructural model, e.g., the neurite orientation dispersion and density imaging (NODDI) model. GRL allows to perform fiber tractography with higher accuracy than dRL, without the need to define FOD thresholds for the fiber tractography procedure but rather uses the derived signal fraction maps to terminate the tractography procedures, delivering high quality termination at the GM/WM interface or at the outer GM surface.

## 2. Theory

In the following sections we introduce the essential notions of the original damped Richardson-Lucy implementation (section 2.1), and show how GRL accommodates multi-shell diffusion MRI data to perform multi-tissue decomposition (section 2.2). Further, in sections 2.3 and 2.4, we show how different WM models and signal representations can be used to generate the deconvolution kernel, as well as how to take into account intra-voxel incoherent motion (IVIM) effects to quantify signal fractions of blood pseudo-diffusion. In this work, we use the word *model* interchangeably to identify both microstructural models and representations of the diffusion signal, such as the diffusion tensor and kurtosis imaging methods (Novikov et al., 2018). While this might introduce some unclarity, both family of methods can serve as basis function for our deconvolution method, making their differentiation irrelevant to this work.

## 2.1 The original dRL spherical deconvolution approach

The Richardson-Lucy approach (Dell'Acqua et al., 2007) derives the FODs directly in the signal domain, which avoids artefacts that can arise from alternative representations as spherical harmonics, including truncation and Gibbs ringing. The Richardson-Lucy framework requires to define a deconvolution kernel *H*, or H-matrix, that projects the diffusion MRI signals measured along each gradient direction onto the unit sphere. Given the signal *s* acquired at a single diffusion weighting (b-value) applied along multiple gradient directions, the FOD of each voxel can be estimated with an iterative expectation maximization approach, as reported in Eq. [1].

$$[FOD^{(k+1)}]_i = [FOD^{(k)}]_i \frac{[H^T s]_i}{[H^T H \cdot FOD^{(k)}]_i} \quad [1]$$

In the equation above, *FOD*$^{(k)}$ is the FOD estimated at the $k^{th}$ iteration along the *i*-th direction sampled on the unit sphere, *H* is the *m* by *n* deconvolution kernel that maps the *m* (e.g., 60) measured signals onto *n* (e.g., 300) arbitrarily discretized directions uniformly sampled on the unit sphere, and *H*$^T$ is the transpose of **H**. Eq. [1] is semi-convergent, thus its solution practically requires a stopping criteria. Common criteria are the definition of a maximum number of iterations (e.g. 200 to 2000), or a pre-defined threshold on the FOD changes between consecutive iterations.

The damped Richardson-Lucy (dRL) approach extends Eq. [1] with an extra regularization term, to attenuate over-fitting of the FOD to noise or isotropic compartments like CSF. The regularized iterative deconvolution of dRL is described by Eq. [2].

$$[FOD^{(k+1)}]_i = [FOD^{(k)}]_i \left(1 + [u^{(k)}]_i \left(\frac{[H^T s - H^T H \cdot FOD^{(k)}]_i}{[H^T H \cdot FOD^{(k)}]_i}\right)\right) \quad [2]$$

In Eq. [2], the regularization term $u^{(k)}$ is updated at each iteration *k* based on the amplitude of the current FOD estimation and the standard deviation of the diffusion weighted signals *std(s)*, as shown in Eq. [3] and Eq. [4].

$$[r^{(k)}]_i = 1 - \frac{[(FOD^{(k)})^v]_i}{[(FOD^{(k)})^v + \eta^v]_i} \quad [3]$$

$$u^{(k)} = 1 - \max(0.1 - 4std(s)) r^{(k)} \quad [4]$$

In this work, the constant $v$ was set to $v$=8 (Dell'Acqua et al., 2010), whereas $\eta$ was set to two times the maximum amplitude of the FOD describing an isotropic diffusion process with diffusion coefficient 0.7x10$^{-3}$mm$^2$/s. Such value of $\eta$ is chosen to minimize the contribution of GM to the FOD of WM.

## 2.2 The GRL approach for multi-shell diffusion MRI data

In section 2.2.1 we describe the GRL framework to describe the diffusion signal from multiple tissue types by using multi-shell dMRI. In sections 2.2.2 to 2.2.5 we show some possible model choices to generate the deconvolution kernel **H** that describes WM.

2.2.1 Multi-shell and multiple tissues in GRL

Taking advantage of the additional contrast provided by multiple diffusion weighting (b-values/shells) it is possible to disentangle different tissue contributions to the diffusion MRI signal. The extension of dRL to support multiple diffusion weighting requires to extend the deconvolution kernel (H-matrix) along the rows, which can be formulated as in Eq. [5].

$$H_{GRL} = \begin{bmatrix} H_{dRL,1} \\ \vdots \\ H_{dRL,p} \end{bmatrix} \quad [5]$$

Here p is the total number of considered diffusion weighting (i.e., shells). In practice, the H-matrix introduced with GRL consists of the stacking of multiple H-matrices corresponding to each of the acquired shells. To account for mixtures of multiple tissue types (partial volume effect), the problem can be formulated by adding further columns to the H-matrix to describe the tissues of interest. In name of simplicity, here we will consider a formulation similar to a previous study introducing multi-shell CSD (Jeurissen et al., 2014). Therefore, we will consider three tissue types when creating the H-matrix, but other choices, e.g. considering intra- vs extra-cellular signals (Alexander et al., 2002), could be also taken into account. Considering the different diffusion characteristics of WM, GM and CSF, their signals can be taken into account in the H-matrix formulation as in Eq. [6].

$$H_{GRL} = \begin{bmatrix} H_{dRL,1,WM} & H_{dRL,1,GM} & H_{dRL,1,CSF} \\ \vdots & \vdots & \vdots \\ H_{dRL,p,WM} & H_{dRL,p,GM} & H_{dRL,p,CSF} \end{bmatrix} \quad [6]$$

In the following sections, we refer to GRL as the formulation with three tissue types representing WM, GM and CSF, if not specified. The columns of H-matrix corresponding to isotropic GM and CSF contributions can be generated with the apparent diffusion coefficient (ADC) equation and using typical values of the two compartments, e.g. $0.7 \times 10^{-3}$ mm$^2$/s and $3.0 \times 10^{-3}$ mm$^2$/s. Regarding the modeling of WM, the following sections discuss some possible model choices to generate the H-matrix. We consider both representations of the diffusion signal such as the diffusion tensor with a single fixed tensor (GRL-FT), the diffusion kurtosis imaging model (GRL-DKI), and a simplified white matter model from the NODDI framework (GRL-NODDI), as explained in the following sections.

2.2.2 DTI-based H-matrix with fixed tensor (GRL-FT)

The original formulation of the single-shell dRL method (Dell'Acqua et al., 2010) used the DTI representation and a predefined set of eigenvalues to initialize the H-matrix. Similarly, we generated the columns of the H-matrix of GRL-FT using eigenvalues equal to $[1.7, 0.2, 0.2] \times 10^{-3}$ mm$^2$/s. This choice of the eigenvalues implicitly assumes non-Gaussian effects to be negligible.

2.2.3 DKI-based H-matrix (GRL-DKI)

The diffusion signal is known to be non-Gaussian at strong diffusion weighting (Jensen et al., 2005) in the brain. Deviations from Gaussian diffusion are typically observed in GM and, to a bigger extent, in WM. Previous studies (Jensen et al., 2005; Jensen & Helpern, 2010) introduced the DKI model, an extension of DTI to account for the restriction of water molecules observed at strong diffusion weighting, as described in Eq. [7].

$$ln(S_{DWI}) = lnS_0 - bD - \frac{1}{6}b^2D^2K \qquad [7]$$

In the above equation, $S_{DWI}$ and $S_0$ are the diffusion weighted and the non-diffusion weighted signals, respectively, *D* is the diffusion coefficient, *b* the diffusion weighting, and *K* the value of mean kurtosis. The columns of the H-matrix of GRL-DKI were generated with Eq. [7] using tensor values and mean kurtosis *K* values estimated from the data. All the data was fit with a DTI model extended to account for isotropic kurtosis (De Luca, Leemans, et al., 2018) and using a weighted linear least squares estimator (Veraart et al., 2013). The values of the tensor and of *K* were averaged within a mask defined by fractional anisotropy values above 0.7.

2.2.4 Microstructure-based H-matrix (GRL-NODDI)

The columns of the H-matrix describing WM can be modelled not only with statistical descriptions of the diffusion signal, such as DTI and DKI, but also with microstructural models. As a proof of concept, here we consider the neurite model defined in the neurite orientation dispersion and density imaging (NODDI) (H. Zhang et al., 2012). We generated the columns of the H-matrix of GRL-NODDI by setting the intracellular volume fraction to 1, the intrinsic free diffusivity of the medium to 1.7 x $10^{-3}$ mm$^2$/s, the concentration parameter of the Watson distribution to 3.5, and the isotropic volume fraction to 0. These values are meant to represent the typical an average NODDI parameter set observed in WM, with the exception of CSF for which we account explicitly in GRL.

**2.3 A two-step iterative fit approach**

The solution of the deconvolution problem with the above-mentioned H-matrices cannot be derived using directly the iterative dRL scheme, because all terms would contribute to the final FOD. For this reason, we decoupled the problem into two steps, namely the estimation of the tissue fractions and the FOD calculation. The new GRL approach can be written as an iterative process alternating the dRL algorithm and a non-negative least-squares estimation, described in Eq. [8] and Eq. [9].

$$\text{FOD}^j = \mathbf{dRL}\left(s - \mathbf{Y}\begin{bmatrix} f_{GM}^{j-1} \\ f_{CSF}^{j-1} \end{bmatrix}\right) \qquad [8]$$

$$\begin{bmatrix} f_{WM}^j \\ f_{GM}^j \\ f_{CSF}^j \end{bmatrix} = \underset{f \geq 0}{argmin} \left\| [\mathbf{H} * \widetilde{\text{FOD}}^j \quad \mathbf{Y}] \begin{bmatrix} f_{WM} \\ f_{GM} \\ f_{CSF} \end{bmatrix} - s \right\|_2^2 \qquad [9]$$

In the above equation, $f_{WM}$, $f_{GM}$ and $f_{CSF}$ represent the signal fractions associated to each compartment at iteration *j*, *s* is the measured signal (normalized by the non-weighted image), **H** is the H-matrix describing the fiber response, and **Y** is an m by 2 matrix containing the signal representation of the isotropic contributions (GM, CSF). The * operator symbolizes a spherical convolution operation to compute the signal given the estimated FOD. In eq. [9] we use an intrinsic regularization function for the FOD function to further penalize spurious peaks, which is the median thresholding of the FOD, denoted as $\widetilde{\text{FOD}}$ in Eq. [9]. The values of fractions $f_{GM}$ and $f_{CSF}$ were set to zero in the first iteration.

The modularity of the GRL framework allows to easily take into account more isotropic compartments by simply expanding the deconvolution matrix along the columns dimension. As a proof of concept, here we show that in addition to WM, GM and CSF, it is also possible to account for blood pseudo-diffusion using the intra-voxel incoherent motion (IVIM) model (Le Bihan et al., 1986, 1988). The diffusion coefficient of the IVIM component was set to an arbitrary value of 50.0 x 10$^{-3}$ mm$^2$/s to generate an additional column of the isotropic deconvolution matrix **Y**. The adjusted fit procedure to account for WM, GM, CSF and IVIM is reported in Eq. [10] and Eq. [11].

$$\text{FOD}^j = \mathbf{dRL}(s - \mathbf{Y}\begin{bmatrix} f_{GM}^{j-1} \\ f_{CSF}^{j-1} \\ f_{IVIM}^{j-1} \end{bmatrix}) \qquad [10]$$

$$\begin{bmatrix} f_{WM}^j \\ f_{GM}^j \\ f_{CSF}^j \\ f_{IVIM}^j \end{bmatrix} = \underset{f \geq 0}{argmin} \left\| [\mathbf{H} * \widetilde{\text{FOD}^j} \quad \mathbf{Y}] \begin{bmatrix} f_{WM} \\ f_{GM} \\ f_{CSF} \\ f_{IVIM} \end{bmatrix} - s \right\|_2^2 \qquad [11]$$

To numerically solve Eq. [10] and [11], the number of acquired shells must be larger than the number of modelled compartments, to make sure the columns of H are not rank-deficient. In the case of Eq. [10] and [11], this means that at least four unique diffusion weightings are needed, of which at least one where the IVIM contribution is not null. An implementation of the abovementioned steps is freely available as part of the toolbox *MRIToolkit* (https://github.com/delucaal/MRIToolkit).

## 3. Methods

Single-shell dRL and GRL were fit to both simulated and in-vivo data. The H-matrix used for dRL was generated in agreement with previous work (Dell'Acqua et al. 2010), e.g., with a diffusion tensor described by the eigenvalues [1.7, 0.2, 0.2] x 10$^{-3}$ mm$^2$/s.

**3.1 Weighting of multi-shell data**

Data acquired at strong diffusion weighting has high angular resolution but generally poor SNR, whereas data at low diffusion weighting has high SNR but poor high angular resolution. In the GRL formulation, the contribution of each shell to the fit can be weighted (Eq. [8-11]) by multiplying both the signal and the rows of the H-matrix corresponding to a specific shell by a scalar ω. In this way, the data at all diffusion weightings is used simultaneously in the deconvolution operation without any averaging, but the weight of the fit residuals of the lower diffusion weighting is scaled with ω during the FOD determination. In this analysis, the value of ω was kept identical for all the inner shells. To investigate the effect of low diffusion weighting data on the angular resolution, we simulated a single-voxel crossing fiber configuration with 60° crossing angle and 1000 realizations of Rician noise at SNR 50. We performed the fit with GRL-FT by varying the weight of all shells except the outer between 0.1 and 1. GRL-FT was chosen as representative because this analysis focuses on the effect of data at low diffusion weightings on the angular resolution, but results are expected to generalize to the other considered H-matrices. Further, in the Supplementary Material we repeated the same analysis for several crossing fiber configurations with angles between 20° and 90°. These simulations were performed without noise, to investigate purely the effect of the multi-shell weight on the angular resolution of GRL as compared to dRL.

**3.2 Simulations**

The robustness to noise of GRL was tested with numerical simulations using the diffusion gradient encoding scheme of the Human Connectome Project. Single fiber populations were generated with a three compartments model accounting for WM, GM and CSF. The DTI model was employed for all three compartments, setting the eigenvalues of the WM tensor equal to [1.7, 0.2, 0.2] x $10^{-3}$ mm$^2$/s, while GM and CSF were considered isotropic with mean diffusivity values of 0.7 x $10^{-3}$ mm$^2$/s and 3.0 x $10^{-3}$ mm$^2$/s, respectively. Three different volume fractions ($f_{WM}$, $f_{GM}$, $f_{CSF}$) mixtures were generated under the constraint of $f_{WM} + f_{GM} + f_{CSF} = 1$, including:

1. $f_{GM}$ = 0, $f_{WM}$ assuming values between 0 and 1 with step 0.01;
2. $f_{CSF}$ = 0, $f_{WM}$ assuming values between 0 and 1 with step 0.01;
3. $f_{WM}$ assuming values between 0 and 1 with step 0.01, $f_{GM}$ = $f_{CSF}$ = 0.5 x (1 - $f_{WM}$).

Realizations of Rician noise were added to the signals to simulate the SNR levels at b = 0 s/mm$^2$ equal to [20, 30, 50, 1000] with 1000 repetitions each. As a measure of performance, the derived signal fractions and the FOD peak orientations were compared to the simulated values.

**3.3 In-vivo datasets**

3.3.1 Data acquisition

Four in-vivo datasets with different acquisition protocols were used in this work to demonstrate the performance of GRL. (1) Data1 is the diffusion MRI data of a subject from the Human Connectome Project (HCP) (Van Essen et al., 2012). It includes 18 volumes acquired at b = 0 s/mm$^2$, and 90 gradient directions at b = 1000, 2000, 3000 s/mm$^2$ with a voxel size of 1.25x1.25x1.25 mm$^3$. In addition to the diffusion data, a T1-weighted scan co-registered to the diffusion data was also available. (2) Data2 is a clinical dataset acquired with a 3T Philips scanner, and includes 13 b = 0 s/mm$^2$ images, 32 volumes at b = 700 s/mm$^2$, 55 volumes at b = 1000 s/mm$^2$, and 100 volumes at b = 2500s/mm$^2$ with a voxel size of 2.0x2.0x2.0 mm$^3$. (3) Data3 is a subset of the MASSIVE dataset (Froeling et al., 2017), a multi-session dataset collected with a 3T scanner at the UMC Utrecht with voxel resolution 2.5x2.5x2.5 mm$^3$. It includes 420 non-weighted volumes, 30 volumes at b = 5, 10, 250 s/mm$^2$, 40 volumes at b = 25, 40 s/mm$^2$, 20 volumes at b = 60, 80, 140 s/mm$^2$, 250 volumes at b = 500 s/mm$^2$ and 500 volumes at b = 1000, 2000, 3000 s/mm$^2$. (4) Data4 is a multi-shell dataset scanned twice (test-retest) within the same session. This dataset includes a T1-weighted scan with resolution 1x1x1 mm$^3$ (echo time TE=3.7ms, repetition time TR=8ms, parallel imaging SENSE 2+2.6, flip angle 8°), and a diffusion MRI scan with 8 volumes at b = 0s/mm$^2$, 10 volumes at b = 700s/mm$^2$, 10 volumes at b = 1200s/mm$^2$ and 45 volumes at b = 2500 s/mm$^2$. The data was acquired at a resolution of 2x2x2.5 mm$^3$, TE=90ms, TR=2s, SENSE factor 2.

3.3.2 Data Analysis

Data2 and Data3 were pre-processed in ExploreDTI (A. Leemans et al., 2009) to attenuate artefacts commonly observed on diffusion MRI data. Both datasets were corrected for signal drift (Vos et al., 2017), then motion and eddy currents correction was performed by means of an affine registration to the first non-weighted image, and the diffusion encoding matrix was rotated accordingly (Alexander Leemans & Jones, 2009). The diffusion MRI acquisition of Data1 was downloaded from the HCP database in its already preprocessed format, which did not require any additional step. The T1-weighted image of Data1 was processed with the pipeline "fsl_anat" of FSL (Jenkinson et al., 2012) to perform skull stripping (S. M. Smith, 2002) and tissue type segmentation (Y. Zhang et al., 2001). Data4 was processed similarly to Data2 and Data3 but included an extra non-linear registration step to the T1-weighted image to correct for EPI distortions. The SNR of all datasets was computed using the non-weighted images by determining the voxel-wise average value of the first 8 b = 0 s/mm$^2$ images divided by their voxel-wise standard deviation, then this was averaged within the brain mask of each subject (A. Leemans et al., 2009).

Data1 and Data2 were fit with dRL and GRL using the above-described H-matrices (section 2.2.2) to evaluate the effect of voxel resolution on the fractional maps. In addition to dRL and GRL, Data1 was fit with the multi-shell constrained spherical deconvolution method (Jeurissen et al., 2014), using the tissue type segmentations derived with "fsl_anat" to initialize the response functions corresponding to WM GM and CSF. The signal fractions estimated with GRL were compared to those derived with MSCSD by computing their Pearson correlation coefficient. Data3 were fit with the with GRL-FT while also considering an additional signal class to account for intra-voxel incoherent motion effects at low diffusion weighting. The WM FOD obtained on the two datasets were compared to those derived with a dRL fit of the largest shell of each dataset. The number of fibers (NuFO) (Dell'Acqua et al., 2013) of each FOD derived with MSCSD, dRL and GRL was derived separately in WM and GM, using the segmentations from the T1-weighted image, then compared between methods. In particular, we counted as "fiber" only peaks with magnitude above 20% the average amplitude of the first peak in pure WM. The amplitude of the first peak for single fiber population WM was derived in voxels with a value of fractional anisotropy above 0.7. Finally deterministic fiber tractography was performed in ExploreDTI with the derived FODs for both dRL, MSCSD and GRL. The tracking procedure was performed on dRL using angle-threshold 45 degrees, step size equal to half of the voxel-size, with seed points evenly sampled in the brain volume per 2 x 2 x 2 mm$^3$. Tractography of the FODs derived with dRL was performed first with an FOD threshold equal to 0.1, which corresponds to the FOD amplitude of a GM voxel and is meant to stop the tractography at the WM/GM interface. Subsequently, the tractography was repeated with an FOD threshold equal to 0.001 to allow the continuation of the tracts into cortical GM. For the tracking of FODs derived with GRL and MSCSD, equal parameters but no FOD threshold were used. The signal fraction maps associated to WM and GM, which are derived with GRL or MSCSD but not with dRL, were used as termination criteria to stop tracking at the GM/WM interface, and at the outer GM surface, respectively, to make a fair comparison. The reliability of GRL was tested with Data4 and compared to MSCSD and dRL. The details and results of this analysis step are reported in the Supplementary Material.

## 4. Results

The effect of including shells at multiple diffusion weightings in the FOD estimation is shown in Fig. 1. Increasing the weight of the shells at b = 1000, 2000 s/mm$^2$ in the fit reduces the angular resolution of data at b = 3000 s/mm$^2$, inducing angular deviations on the first peak up to 4.2°. Including data at multiple diffusion weightings is, however, essential to account for partial volume effects, which can be optimally estimated at $\omega = 0.2$ and results in a partial volume estimation error of $f_{GM}$ equal to 3±12%, and a corresponding angular deviation of 1.2°, as shown in Supplementary Material Figure S1. This weighting value was therefore used throughout this manuscript for all datasets. Additional simulations for multiple crossing fiber configurations are shown in Supplementary Material Figure S2. The figure shows that values of $\omega$ above 0.2 worsen the angular resolution of GRL-FT for crossing angles below 60°. Further, the choice of the eigenvalues used in GRL-FT determines the minimum crossing angle that can be resolved, which is 50° for the settings used throughout this manuscript but can be lowered to up to 37° with other choices of the H-matrix parameters.

The signal fraction maps estimated with GRL-FT in Simulation I-II-III and their standard deviations are shown in Fig. 2. At SNR 20, the precision of the estimation of small CSF-like partial volume effects drops ($f_{WM}$ > 0.9, $f_{CSF}$ < 0.1). For SNR levels above 20, GRL can effectively disentangle partial volume effects with high accuracy. For linearly decreasing partial volume effects, linearly increasing values of $f_{WM}$ are estimated, although a small but consistent overestimation of $f_{WM}$ – underestimation of $f_{GM}$ – of about 10% is revealed.

The average angular error of the main FOD orientation estimated with dRL and GRL-FT in Simulations I-II-III are shown in Fig. 3. While dRL can effectively suppress the effect of partial volume effects on the FOD for $f_{WM}$ values above 0.5, angle errors up to 50 degrees are observed for lower $f_{WM}$ values, in particular at SNR 30. Conversely, FODs estimated with GRL have less angular error in presence of strong partial volume effects for SNR above 20, with a maximum angular error of 9 degrees for $f_{WM}$ = 0.2. At lower SNR levels (SNR 20), the application of GRL-FT is still beneficial in comparison to dRL, but FODs are generally poorly resolved when the WM content drops below 50%. For high SNR levels (SNR $\geq$ 50), dRL results in minimal angular error for $f_{WM} \geq 0.2$.

The average SNR of Data1, Data2 and Data3 were 21, 24 and 24, respectively. Fig. 4 shows an example axial slice of the signal fraction maps estimated with GRL with Data1 and Data2, and with the T1-weighted segmentation and MSCSD on Data1. The values of $f_{WM}$, $f_{GM}$ and $f_{CSF}$ are plausible, in line with the expected anatomy, and consistent between the datasets despite differences in imaging resolution and diffusion gradient encoding. The deep GM was characterized by high values of the GM fraction, with the thalamus, caudate nucleus and lentiform nucleus clearly visible in the GM fraction map of both Data1 and Data2. The choice of the H-matrix affects the estimated signal fraction maps. With GRL-FT and GRL-DKI the $f_{GM}$ map assumes zero-values within white matter, whereas GRL-NODDI produce non-zero $f_{GM}$ values in such region. Further, GRL-NODDI results in remarkable higher $f_{CSF}$ values as compared to the other three H-matrix choices. When comparing the signal fractions estimated with GRL to those from MSCSD and the T1-weighted segmentation, remarkable similarity can be observed. In GM, the Pearson correlation coefficients between $f_{GM}$ derived with MSCSD and that computed with GRL were 0.81 for GRL-FT, 0.88 for GRL-DKI and 0.70 for GRL-NODDI. The Pearson correlation coefficients between $f_{WM}$ computed with MSCSD in WM and that computed with GRL were 0.92 for GRL-FT, 0.94 for GRL-DKI and 0.96 for GRL-NODDI.

GRL can be used to take into account more than three tissue models, including the IVIM effect observed at low diffusion weightings. This is shown in Fig. 5 for an axial slice of the MASSIVE dataset. In addition to signal fraction maps associated to WM, GM and CSF remarkably similar to those observed with Data1 and Data2 (Fig. 4), GRL estimated an $f_{IVIM}$ map showing high perfusion values in the middle temporal lobe, and in locations compatible with the middle cerebral artery. The estimated values of $f_{IVIM}$ in tissues is 11±6%.

The FODs estimated with MSCSD (Data1 only), dRL and GRL with Data1 and Data2 in an example coronal slice are shown in Fig. 6 and Fig. 7, respectively. Generally, GRL-FT and GRL-DKI perform similarly to dRL and MSCSD within WM, providing excellent separation of up to three crossing fiber configurations (red box). In proximity of the ventricles (yellow box), GRL-DKI visibly reduces the estimation uncertainty of the FODs, and further suppresses FODs clearly belonging to the CSF area. GRL-NODDI also reduces the effect of partial volume effects on the WM FOD as compared to dRL but is less effective at separating crossing fiber configurations. In the WM of Data1, the average angle between the first peak estimated with MSCSD and that estimated with GRL in WM assumes values between 5.4° ± 6.5° for GRL-DKI and 6.7° ± 6.2° for GRL-FT, as compared to 7.1° ± 6.5° for dRL. The angle deviation of the first peak computed with GRL on Data1 as compared to that computed with MSCSD in GM assumes values between 8.8° ± 9.6° for GRL-DKI and 10.3° ± 12.8° for GRL-NODDI, as compared to 10.9° ± 9.9° for dRL. Similar results were also observed in Data4, as shown in Supplementary Material Table S1. The angular deviation between MSCSD and GRL was within 8° in WM and 9° in GM, and in the same order of the angle deviation between MSCSD and dRL.

Fig. 8 and Fig. 9 show an example axial slice of the FODs estimated with MSCSD (Data1 only), dRL and GRL with Data1 and Data2 in a region where the WM enters the cortical folding (red box). Within WM, the FODs estimated with dRL, GRL-FT and GRL-DKI show remarkable similarity. On Data1, both GRL-FT and GRL-DKI estimate sharper FODs with a reduced number of peaks in comparison to dRL and MSCSD, particularly in proximity of GM/CSF (yellow box). As previously observed, the FODs estimated with GRL-NODDI contain a lower number of crossing fibers also in this region. On Data 2, which has lower spatial resolution than Data1 and includes a diffusion weighting below b = 1000 s/mm$^2$, the FODs estimated with GRL in proximity of the cortex are strongly attenuated as compared to those estimated with dRL.

To investigate to which extent GRL resulted in a different number of spatial orientations as compared to dRL, we derived the FOD peaks of both methods throughout the brain which amplitude was above 20% that of a typical WM peak. Results, which are reported in Fig. 10, show that GRL-FT performs very similarly to dRL with the exception of a slight reduced number of voxels with 3 or more crossing fiber configurations, which visually are located in proximity of the cortex. GRL-DKI results in a reduced number of three fiber crossings throughout the WM as compared to GRL-FT but shows very similar results to MSCSD. Nevertheless, this result might be driven by the peak thresholding as such fiber configurations can be appreciated in the FODs shown in Fig. 6 and Fig. 7. In GM, GRL-DKI and GRL-NODDI essentially reduced the number of observed peaks to 1.

The whole brain fiber tractography of Data1 and Data2 are shown in Fig. 11 and Fig. 12. The figure shows the fiber tractography results obtained with dRL for a threshold equal to the average FOD amplitude in GM (t=0.1) – which should therefore terminate the tracking at the WM/GM interface – and without using constraints on the FOD amplitude (t=0.001), allowing the tracking to continue to the cortical surface (and beyond). When tracking with GRL and MSCSD, we did not use any FOD-based termination criteria but rather stopped the tracking when $f_{WM}$=0 and $f_{GM}$=0, respectively. On both datasets, we observe that GRL provides tractography results that are practically identical to those provided by dRL within WM, as exemplified by the red tracts shown in Fig. 11 and Fig. 12. Tractography with MSCSD and GRL-DKI shown most comparable results when tracking goes into WM/GM interface. On Data2, however, the WM tractography in the cortex provided by dRL showed some spurious – likely artefactual – fiber reconstructions in the frontal lobe, which were not observed with GRL. When removing the FOD tractography threshold, tractography results provided by dRL (t=0.001) continued erroneously outside the cortical grey matter. On the contrary, using the grey matter mask with GRL to constrained fiber tractography shows much cleaner results with a physiologically plausible delineation of the cortical ribbon, and a generally plausible continuation of WM bundles into the cortex (yellow box).

The reliability analysis of MSCSD, dRL, GRL in a test-retest experiment is reported in the Supplementary Material. The inter-scan angular deviation obtained with GRL was in the same order of that observed with dRL and MSCSD, as reported in Supplementary Material Table S1. When performing fiber tractography of the corticospinal tract and comparing to MSCSD, the highest DICE score was observed for GRL-DKI (0.80). Further, GRL-DKI achieved the highest inter-scan DICE (0.86), performing slightly better than dRL (0.85) and MSCSD (0.81), as shown in Supplementary Material Table S2.

## 5. Discussion

In this study, we generalize the Richardson-Lucy framework (GRL) to leverage the information content of multi-shell diffusion MRI data and model multiple tissue types. GRL improves the quality of the FOD estimation step by fitting dMRI models of choice – including DTI, DKI and NODDI – to data acquired at multiple diffusion weightings to disentangle partial volume effects of WM with GM, CSF and other intra-voxel incoherent motion effects (IVIM). In addition to reducing the effect of partial volume effects on WM FODs, GRL uses the signal fraction estimates to terminate the fiber tractography step, ultimately reducing spurious reconstructions without the need to explicitly define threshold-based stopping criteria, similarly to the ACT framework (Girard et al., 2014; R. E. Smith et al., 2012).

Previously, the study of Dell'Acqua (Dell'Acqua et al., 2010) introduced the dRL framework to reduce the number of spurious peaks in the FOD due to partial volume effects, showing the importance of accounting for isotropic contaminations of cerebrospinal fluid in spherical deconvolution. More recently, Jeurissen (Jeurissen et al., 2014) have shown that by using multi-shell diffusion MRI data and modeling explicitly the contributions of CSF and GM in the spherical deconvolution, it is possible to effectively improve the FOD estimation accuracy as compared to deconvolution methods using only single-shell diffusion MRI data (Roine et al., 2014). In analogy with the work of Jeurissen (Jeurissen et al., 2014) in the context of constrained spherical deconvolution, here we revisit the dRL approach to improve the accuracy of the FOD estimation by taking into account multiple tissue types and the information content of multi-shell diffusion MRI data. Differently from other approaches, and in particular from the work of MSCSD (Jeurissen et al., 2014), GRL performs the spherical deconvolution operation in the original signal space rather than in the spherical harmonics basis (Cheng et al., 2014; Christiaens et al., 2017; Tournier et al., 2004), and uses a model-based response function rather than estimating it from the data (Tax et al., 2014; Tournier et al., 2007). While a more extensive comparison between the two methods can be found elsewhere (Dell'Acqua & Tournier, 2018; Parker et al., 2013), there are several advantages to apply the GRL framework. 1) GRL is compatible with multi-shell data, and more generally with diffusion acquisition schemes featuring multiple diffusion weightings, such as data acquired with non-spherically distributed diffusion gradients, without performing any q-space interpolation (Wedeen et al., 2005). GRL does not require data sampled on a q-space grid as in diffusion spectrum imaging, and is compatible with CHARMED-like acquisitions where 3-6 gradient directions are applied at increasing diffusion weighting over a wide range of diffusion weightings. 2) The choice of using a model-based approach in GRL potentially allows to apply the method to tissues where the response function is difficult to objectively isolate but a signal representation is available, such as the kidneys (De Luca, Froeling, et al., 2018). It is straightforward to account for multiple tissue classes in the GRL framework with models of choice, including the shown pseudo-diffusion and free-water signals. 3) GRL explicitly uses the computed signal fractions to terminate fiber tractography, removing the need to set parameters as the FOD threshold, which are difficult to optimize objectively.

The SNR of dMRI data typically degrades with the strength of the applied diffusion gradient, which conversely enhances the angular resolution of the data (Tournier et al., 2013). Performing multi-shell deconvolution might allow to combine the benefits of both, but this has not been previously well investigated. In practice, it appears beneficial to differently weight data at different diffusion weightings during the spherical deconvolution procedure, to avoid degrading its angular resolution. In Fig. 1, we investigated a new way to include data at lower diffusion weighting in the spherical deconvolution without worsening the angular resolution. We established that weighting the intermediate shells at about 20% represents an optimal choice in our study, trading off angular resolution in exchange of the ability to model tissue types over different diffusion weightings, a required step to allow for signal formulations as NODDI (H. Zhang et al., 2012). In practice, small values of the multi-shell weighting are preferable to prevent angular resolution degradation, and its effect is more prominent when using as H-matrix a diffusion tensor with FA = 0.7 than an impermeable cylinder, as shown in Supplementary Material Figure S2. It should be noted, however, that this choice of the weighting represents an optimal scenario for Data1, and that the weighting optimization should be repeated in sight of the effectively considered gradient scheme. Further, the suggested scheme weights rather simplistically all the lower diffusion weightings with a fixed factor to demonstrate how this affects the angular performances of spherical deconvolution, but schemes uniquely weighting data at each diffusion weighting might further improve the performances of GRL.

With simulations, we investigated the performances of GRL at disentangling partial volume effects of WM-like, GM-like and CSF-like compartments at different SNR levels. The results shown in Fig. 2 indicate that GRL can estimate three class mixtures with high accuracy when the SNR is equal or above 30. A small but consistent bias is observed that leads to overestimation of $f_{WM}$ and underestimation of $f_{GM}$. Nevertheless, the linear relation between simulated and estimated $f_{WM}$ values is not affected by such effect, which we deem negligible to the final application of GRL. At SNR 20, the uncertainty on the estimation of small $f_{CSF}$ becomes relatively large, suggesting that partial volume effects below 5% might be undetected if a sufficient SNR level is not met. This is unsurprising, as the accurate and reliable estimation of mixture of exponentials with little signal fraction is a well-known problem in diffusion MRI (De Luca, Leemans, et al., 2018; Pasternak et al., 2009; Pierpaoli & Jones, 2004). Dell'Acqua et al. (Dell'Acqua et al., 2010) introduced the dRL framework for single shell data to reduce the number of spurious peaks in the FOD due to partial volume effects, showing the importance of accounting for isotropic contaminations of cerebrospinal fluid in spherical deconvolution.

Fig. 3 shows that GRL can substantially increase the robustness of the FOD estimation to partial volume effects in comparison to dRL. For very low WM-like signal fractions, e.g. $f_{WM} \leq 0.2$, the angular deviation of the FOD is reduced of a factor about 2 as compared to dRL, which might allow future applications outside pure white matter. At SNR 20, the application of GRL still improves the angular accuracy of the FOD in comparison to dRL, but in GM-like regions the angular error remains relatively large.

Results on in-vivo MRI data suggest that GRL can be applied to datasets characterized by different diffusion weightings and imaging resolution, computing high quality FODs on data from the HCP project (Data1), a dataset acquired with a 3T clinical scanner at a lower spatial resolution (Data2), and the MASSIVE dataset (Data3). The signal fractions estimated with GRL on Data1 and Data2, shown in Fig. 4, are anatomically plausible and show high correspondence to the tissue types ideally included in the modelling, and similarly to what was previously shown with multi-shell CSD (Jeurissen et al., 2014). Indeed, when comparing the signal fractions estimated with GRL to those estimated with MSCSD, we observed Pearson correlation coefficients between 0.8 ($f_{GM}$) and 0.96 ($f_{WM}$), suggesting that the performance of GRL is in-line with current state-of-the-art methods. The choice of the H-matrix has a critical impact on the estimated signal fractions. In this work we have shown the applicability of GRL in combination with common models, such as the diffusion tensor imaging (Basser et al., 1994), the diffusion kurtosis imaging (Jensen et al., 2005) and the neurite orientation dispersion and density imaging models (H. Zhang et al., 2012). The purpose of considering multiple models in our analysis is to enforce the generalizability of the GRL framework, showing its application with both statistical representation of the diffusion MRI signal as well as with microstructural models. With recent work showing the inhomogeneity of the WM microstructure (Schilling et al., 2019), GRL might serve as a supporting framework to perform spherical deconvolution while accounting for microstructure specific models, eventually informed by acquisitions decoding additional details on the tissue microstructure, such as neurite density (Nilsson et al., 2018; Westin et al., 2014).

In our results, the CSF signal fraction maps estimated using the NODDI based H-matrix (GRL-NODDI) have higher values throughout the brain as compared to other H-matrix choices, which suggests that the parameters used to initialize the NODDI models might need further optimization. The number of isotropic components to be considered in brain tissue modelling remains debated. In the context of spherical deconvolution, it has become common to consider WM, GM and CSF separately to improve the performance of FOD estimation. Outside the context of spherical deconvolution, however, the quantification of signal fraction maps associated to different water compartments is of high interest. The CSF fraction map quantified by GRL (and other previous approaches) can be compared to the free water mapping technique (Pasternak et al., 2009), which has been shown to provide additional insights as compared to DTI in neurodegenerative diseases as Parkinson's disease. With diffusion MRI data acquired at low diffusion weighting, it is also possible to quantify pseudo-diffusion signal fractions, which provide valuable information in diseases as brain tumors (Gong et al., 2018). Fig. 5 shows that, if data at sufficient diffusion weightings is available, GRL can suppress partial volume effects on the FOD while quantifying signal fraction maps of WM, GM, CSF and IVIM (Fig. 5). These signal fractions, which are in line with previous reports (Hare et al., 2017), pose a first step to bridge the gap between quantitative multi-compartment models and multi-shell spherical deconvolution. The values of $f_{IVIM}$ estimated in this work, 11±6% are relatively high as compared to previous literature on pseudo-diffusion, and probably reflect an under calibration of the pseudo-diffusion coefficient used in the modelling.

Fig. 6 and Fig. 7 show that GRL estimates high quality FODs in WM, suggesting that the state-of-the-art angular performance of dRL is preserved in this new formulation. When partial volume effects become prominent, however, as in proximity of the ventricles or in cortical regions (Fig. 8, Fig. 9), GRL improves the FOD estimation by reducing the number of spurious peaks as compared to dRL, which results in sharper – more plausible – fiber orientations. Compared to MSCSD, the FODs estimated with GRL-FT and GRL-DKI are sharper and seem more robust to partial volume effects (blue arrow in Fig. 8). As expected, the choice of the model (H-matrix) used to represent WM has a major impact on the estimated FOD. We observe that GRL-FT and GRL-DKI preserve the ability of dRL to efficiently resolve WM crossing fibers, although the results shown in Fig. 10 suggest that GRL-DKI might tradeoff the ability to disentangle a third peak for robustness. However, it should be noted that in Fig. 10 we used an arbitrary threshold on the WM FOD equal to 20% the average amplitude of the first FOD peak in WM. Lowering this threshold might reveal a large number of 3+ crossing fibers in GRL-DKI. In contrast to the above-mentioned, the use of GRL-NODDI worsened the spherical deconvolution performances in comparison to dRL at resolving crossing fibers. Regarding GRL-NODDI, the NODDI model takes into account the effect of axonal dispersion (H. Zhang et al., 2012), which might reduce the angular resolution in the deconvolution context. Optimizing the initialization parameters of GRL-NODDI, eventually on a voxel basis, might improve the FOD estimation while also allowing to estimate the underlying fiber dispersion. When comparing the angle between the main FOD peak estimated with MSCSD to those estimated with GRL, an average baseline difference of about 6° is observed, both on Data1 and on Data4 (Supplementary Material Table S2). Given that very similar differences can be observed also between dRL and MSCSD, it seems plausible that such small differences are due to the different mathematical formulation between the CSD and the dRL framework. dRL achieves comparable angular estimations with only the high b-value shell datapoints, in contrast to GRL and MSCSD that included the inner shells in the FOD estimation. Further, the observed inter-method angular differences are smaller than the intra-method differences computed in a test-retest experiment (Table S2), suggesting that the angular performances of MSCSD, dRL and GRL are – on average – comparable and sufficiently reliable for in-vivo application. Besides the angular resolution, identifying the correct number of peaks is critical when computing measures such as the apparent fiber density (David Raffelt et al., 2012) or the number of fiber orientations (Dell'Acqua et al., 2013), which are becoming used in clinical studies (Genc et al., 2018). Among the four tested H-matrices, GRL-DKI seems to provide the best choice (Fig. 10), providing good performances in the solution of the fiber crossing problem while improving the fit quality by accounting for the non-Gaussian behavior of the signal at stronger diffusion weighting (Jensen et al., 2005). It should be noted that the DKI formulation used in this work only describes an average kurtosis effect, using one single parameter to model the mean kurtosis (De Luca et al., 2017). Extending this formulation to account for the full diffusion tensor might further enhance the performance of DKI-based deconvolution, as the kurtosis tensor has been previously shown able to model crossing fiber configurations (Glenn et al., 2015).

Fiber tractography is one of the most common applications of spherical deconvolution methods. While many tractography methods exist, in this study we focused on deterministic fiber tractography because it emphasizes the property of the reconstructed FODs without introducing additional variability due to probabilistic sampling (Behrens et al., 2007). Once the FOD has been estimated, fiber tractography generally requires defining some termination criteria, including a threshold on the amplitude of the FODs. If the study interest is to track purely the white matter, defining an FOD threshold amplitude equal to that of GM (t=0.1 in this study) can provide an effective solution, as shown for the tractography results of dRL FODs in Fig. 11 and Fig. 12. Nevertheless, it should be noted that such approach might also lead to the removal of genuine WM peaks, and might be prone to inhomogeneity in the image, due to either B1 or receive field performance. The problem worsens when attempting to study tracts entering in GM, as commonly done in connectivity studies. To allow tracts entering the cortex, the FOD threshold should be lowered, but no explicit criteria can be easily established onto which threshold value to use. Removing the FOD threshold termination criteria is generally also not an option, as it would generate a large number of false positives, as shown for dRL in Fig. 11 and Fig. 12 (condition t=0.001). The GRL framework offers an elegant solution, as it allows to disregard FOD thresholding, and use the information of the signal fraction maps to constrain the tracking procedure. As shown in Fig. 11 and Fig. 12, following this approach tractography can be stopped (approximately) at the GM/WM interface by using the WM signal fraction map as termination criteria, or allowed to enter the cortex by using the GM signal fraction map. Reducing the number of spurious tracts can potentially contribute to reduce the false positive problem, which has been shown being critical in current fiber tractography (Maier-Hein et al., 2019). Furthermore, by improving the accuracy of WM fiber tracking in proximity of the cortex, GRL can potentially provide better results than dRL when computing the structural connectome (Tau & Peterson, 2009). Indeed, in connectomics, graph representations of the brain are constructed by assigning each tract to the closest cortical end-point by dilation of the cortical region, and inaccurate tracking in the GM/WM interface is likely to affect the process (Wei et al., 2017). Another desired characteristic when performing fiber tractography is to achieve consistent / reliable results. With Data4 we investigated how the fiber tractography of the corticospinal tract changes in a scan-rescan experiment with MSCSD, dRL and GRL. The inter-method DICE score between the tracts computed with MSCSD and those computed with dRL and GRL was reasonable, with values around 0.8, suggesting once more the existence of some baseline differences between the results provided by CSD-based and those provided by dRL-based methods. When looking at the intra-method inter-scan DICE score, GRL achieved scores between 0.81 and 0.86, which is equal or better than the DICE score achieved with MSCSD. When performing fiber tractography with dRL, DICE scores comparable to those of GRL were observed, but the value of the chosen FOD threshold (which in GRL is not needed) played a critical role on the inter-scan dice, which varied between 0.69 and 0.85. One reason of more spurious fibers in dRL estimation, especially when the tracking threshold is low (t=0.001), is the relative fewer gradient sampling points resulted from using only single shell data. GRL framework included inner shells, therefore enables more data points to be considered in the FOD estimation. With the same data points and acquisition schemes, GRL show less spurious fibers than MSCSD in the fiber tracking of CST, shown in Figure S3.

Beyond the connectome, fiber tractography is commonly applied in other fields such as neurology (Reijmer et al., 2013, 2017) or surgical planning (Essayed et al., 2017). Brain tumors, among other conditions, can cause presence of edema in the surrounding healthy tissues, which can worsen the performances of fiber tracking unless properly considered (Gong et al., 2018). Similar considerations apply also to conditions such as Parkinson (Ofori et al., 2017) or psychosis (Lyall et al., 2019), where increases of free-water fractions in WM have been reported. We here speculate that another isotropic contamination effect, e.g. the intra-voxel incoherent motion of blood in the micro-vascular network (Le Bihan et al., 1988), might act as an additional confounder in spherical deconvolution along with free water. While the extent of such effect – if any – has not yet been proved, in this work we have shown that it is possible to take it into consideration in the GRL framework (Fig. 5).

Some limitations of our work should be acknowledged. All the test dataset used in this study are brain data. While the brain is the most common target application of spherical deconvolution techniques, the acquisition of multi-shell diffusion MRI protocols outside the brain is feasible, e.g. to resolve crossing fibers in the human tongue muscles (Voskuilen et al., 2019) or in the kidneys (De Luca, Leemans, et al., 2018). GRL is in principle applicable to such districts by considering an appropriate tissue model, but its feasibility remains to be proven. Regarding the application to the brain, recent works suggest to account for three tissue types (Dhollander et al., 2016; Jeurissen et al., 2014) as also considered in this work. However, differently from the mentioned studies we did not estimate the signal representation of each tissue type by using tissue masks, but rather used models with plausible values to represent the considered tissues. Gray matter and cerebrospinal fluid were modeled using the ADC equation and typical representative values taken from literature (Chiapponi et al., 2013; Pasternak et al., 2009). The use of literature-based values has the advantage not to require any additional data and processing beyond the diffusion dataset, standardizing the analysis, but can be inaccurate if any change in the diffusivity of the two tissues would occur. While this was not taken into account in this study, it is straightforward to estimate the diffusivity values in tissue specific masks prior to applying GRL, in analogy with what is done in other methods (Jeurissen et al., 2014). With regards to the modeling of WM, the parameters of the DKI based (GRL-DKI) H-matrices were estimated from the data itself using a WM mask with coherent WM (fractional anisotropy > 0.7), which might be sub-optimal to select only single population WM bundles. Similarly, the parameters used for the single-shell DTI (GRL-FT) and the NODDI (GRL-NODDI) based H-matrices were assumed from previous studies (Dell'Acqua et al., 2007; H. Zhang et al., 2012), and might be sub-optimal when applied to the considered datasets. Nevertheless, we have here shown that GRL allows to perform spherical deconvolution while allowing large freedom to the user in terms of model choice, which parameters can be further optimized in specific applications.

In summary, GRL improves the performance of the Richardson-Lucy spherical deconvolution scheme while also estimating signal fraction maps of multiple water compartments. The derived fractional maps can be used as surrogate to existing techniques for multi-compartment quantification, as free water elimination (Berlot et al., 2014; Henriques et al., 2017; Pierpaoli & Jones, 2004), which have been shown promising in previous clinical studies (Coelho et al., 2018; Hoy et al., 2017). Further, the estimated signal fraction maps of WM and GM can be used as termination criteria for fiber tractography, relieving from the definition of arbitrary thresholds on the fiber orientation distributions.

## 6. Conclusions

In this work we presented the GRL framework, a method to perform spherical deconvolution of multi-shell diffusion MRI data while accounting for partial volume effects to provide better estimation of fiber orientation distributions in white matter. We have shown that GRL can disentangle signal fractions from multiple tissue components, such as white matter, grey matter, cerebrospinal fluid and, if sufficient data is acquired, pseudo-diffusion. The presented framework allows flexibility in terms of acquisition protocols and can accommodate different choices of diffusion models to perform the spherical deconvolution, making it potentially applicable to any tissue for which a signal model is available.

## Acknowledgements

The research of F.G was supported by China Scholarship Council (CSC). The research of A.L is supported by VIDI Grant 639.072.411 from the Netherlands Organization for Scientific Research (NWO). F.D.A acknowledges funding from the Sackler Institute for Translational Neurodevelopment, Institute of Psychiatry, Psychology and Neuroscience, King's College London. A.D.L is supported by ERA-NET NEURON project 4195 - RepImpact.

Figures and captions

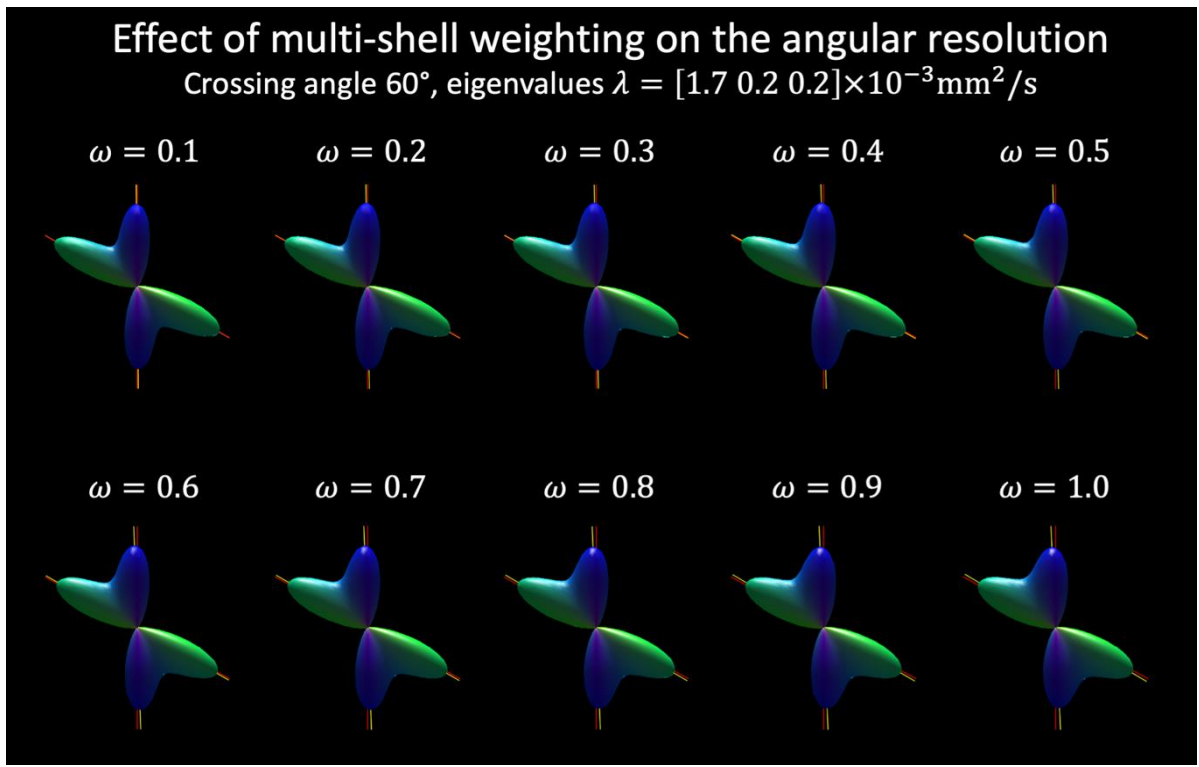

Figure 1.

The effect of multi-shell weighting on a simulated fiber configuration including GM-like partial volume (20%) and 2 WM-like fibers crossing with an angle of 60 degrees (red lines). GRL-FT was applied 1000 times for different noise realizations at SNR 50, then the main peaks were extracted (yellow lines). The angular error increased with increasing weighting of the lower diffusion-weighted shells, but weighting values below $\omega = 0.2$ did not allow to effectively separate partial volume effects. The effect of the multi-shell weighting on the angular error and on the estimated signal fractions is further illustrated in Supplementary Material Figure S1 and Figure S2.

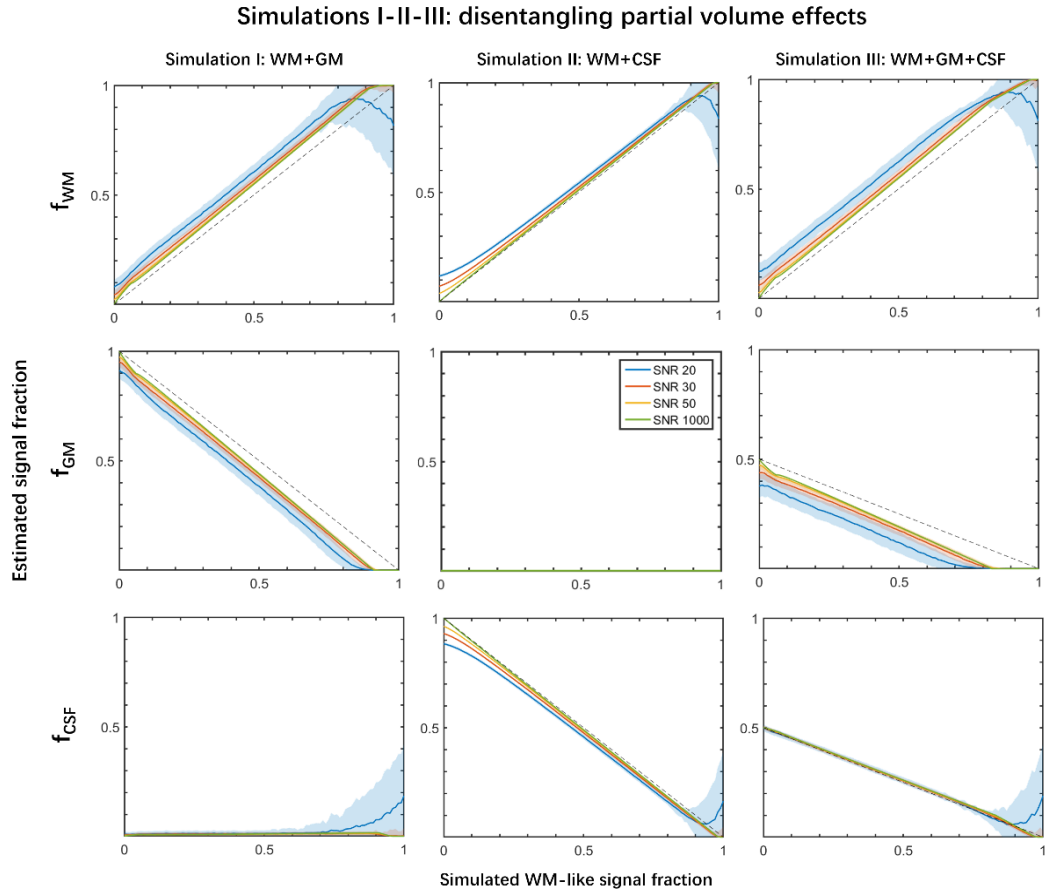

Figure 2.

The mean (solid lines) and the standard deviation (the shadowed area) of the signal fractions estimated with GRL-FT in simulations. The WM-like signal fraction was increased from 0 to 1 with a step of 0.01 and mixed with GM-like signal (Simulation I), CSF-like signal (Simulation II), and both GM and CSF-like signals (Simulation III). GRL-FT recovered the simulated signal mixtures with minimal bias for all SNR levels, a small but consistent overestimation of the WM signal fraction ($f_{WM}$) and underestimation of the GM signal fraction ($f_{GM}$) was observed. The dotted line plots the simulated signal fractions for the corresponding tissue.

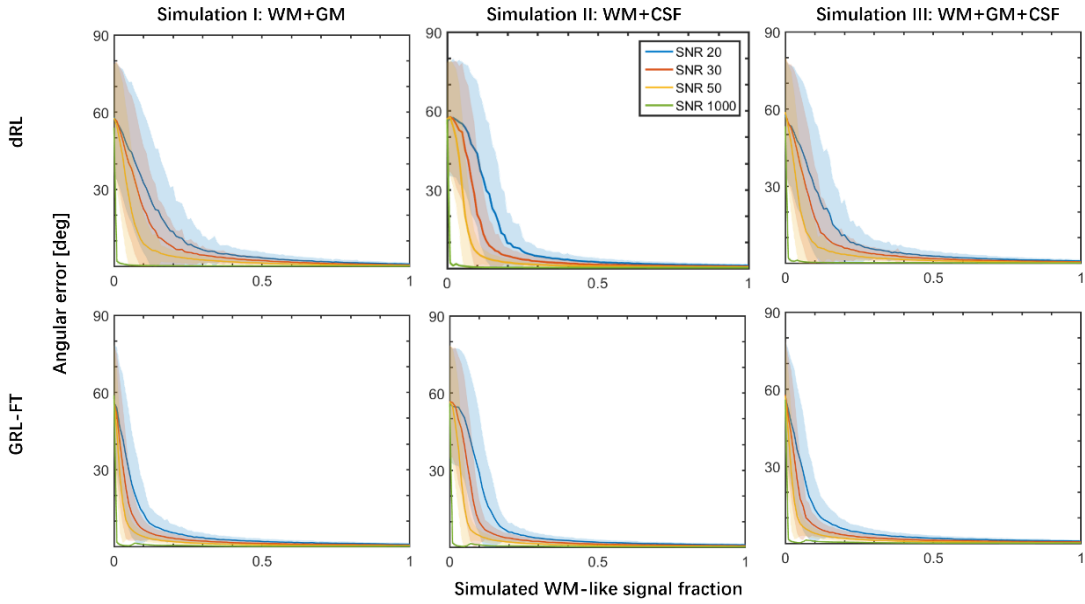

Figure 3.

The mean (solid line) and the standard deviation (shaded area) of the angular deviation of the FODs estimated with dRL (first row) and GRL-FT (second row) in correspondence of multiple SNR levels. WM-like signal was mixed with decreasing GM-like partial volume (Simulation I, first column), decreasing CSF-like partial volume (Simulation II, middle column), and both effects simultaneously (Simulation III, last column). If sufficient SNR is provided (e.g. SNR above 20), GRL-FT can remarkably reduce the angular error as compared to dRL in presence of strong partial volume effects.

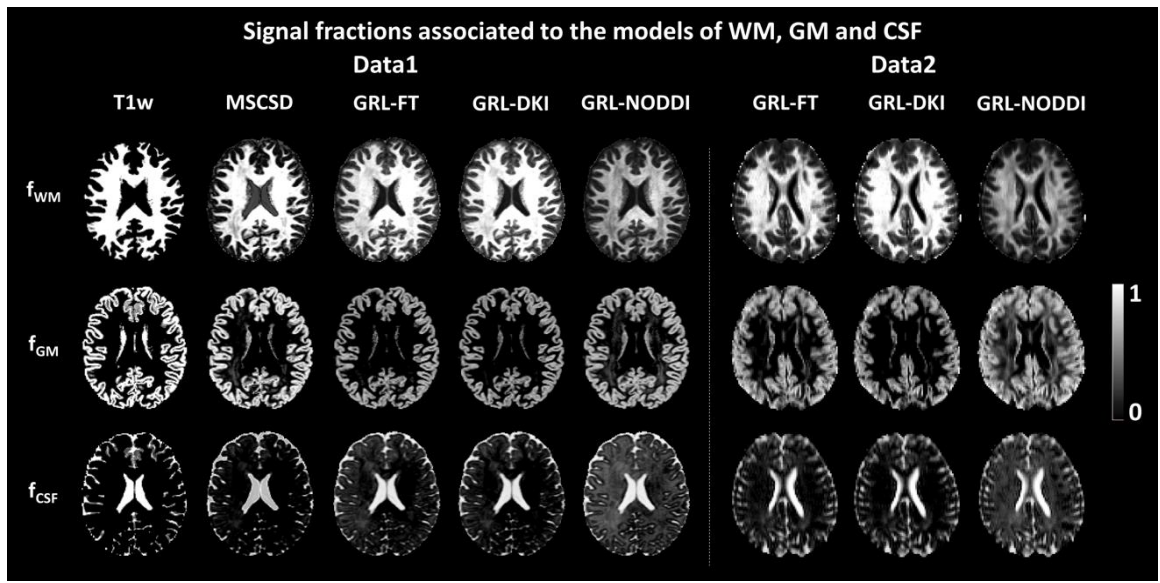

Figure 4.

An example axial slice of the signal fractions estimated with GRL (for different H-matrix choices) on Data1 and Data2. For Data1, the signal fractions estimated from the T1-weighted image (T1w) and multi-shell CSD (MSCSD) are also shown for visual comparison. The signal fractions of WM, GM and CSF estimated with GRL are in line with known anatomy and are spatially homogeneous on both datasets, suggesting numerical stability and voxel-wise specificity of the solution. Further, the signal fractions estimated with GRL show remarkable similarity to those derived with both MSCSD and the T1w segmentation.

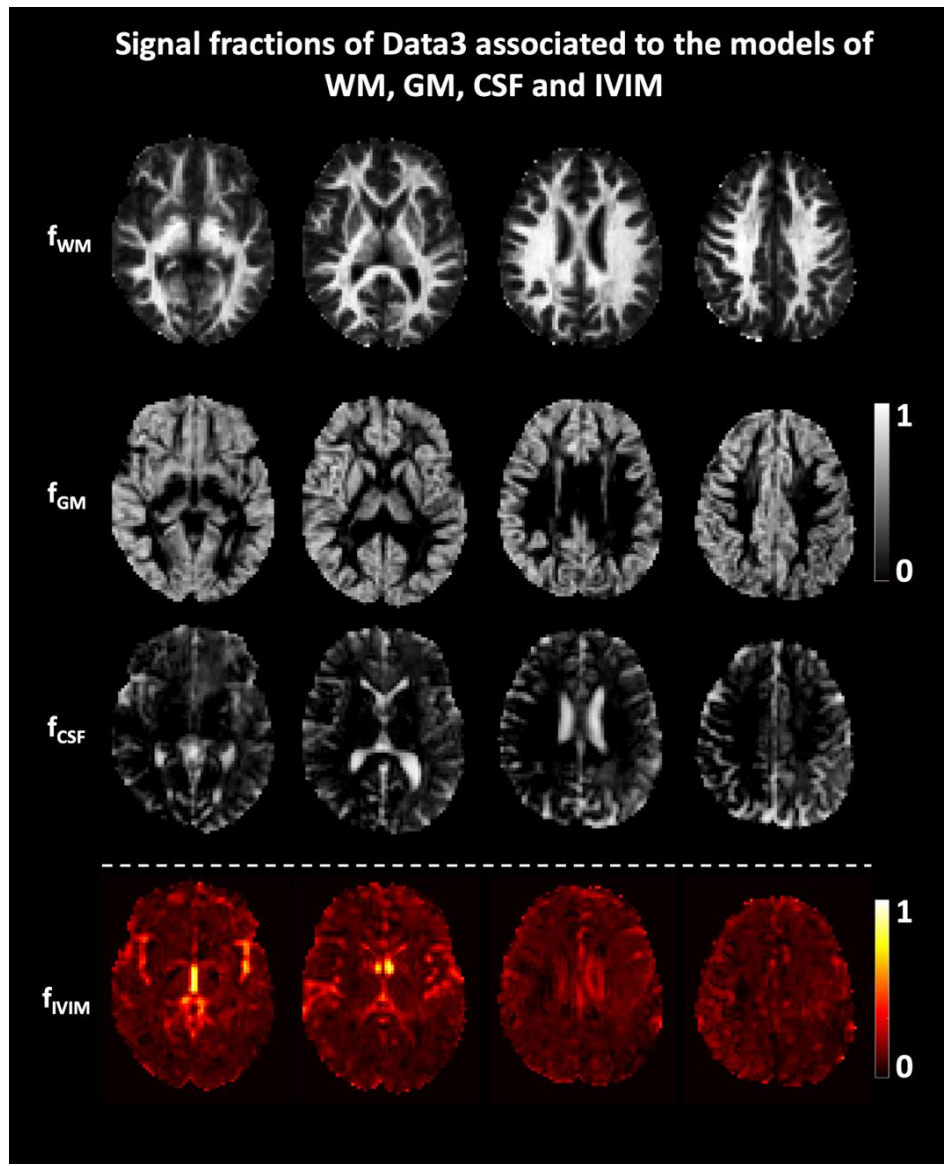

Figure 5.

An example axial slice of the signal fractions estimated with GRL-FT on Data3 when also including pseudo-diffusion effects into the H-matrix formulation to quantify IVIM effects in addition to WM, GM and CSF. The estimated IVIM contributions are non-zero throughout the brain, and are in-line with known anatomy, with signal fraction values around 5% in tissues.

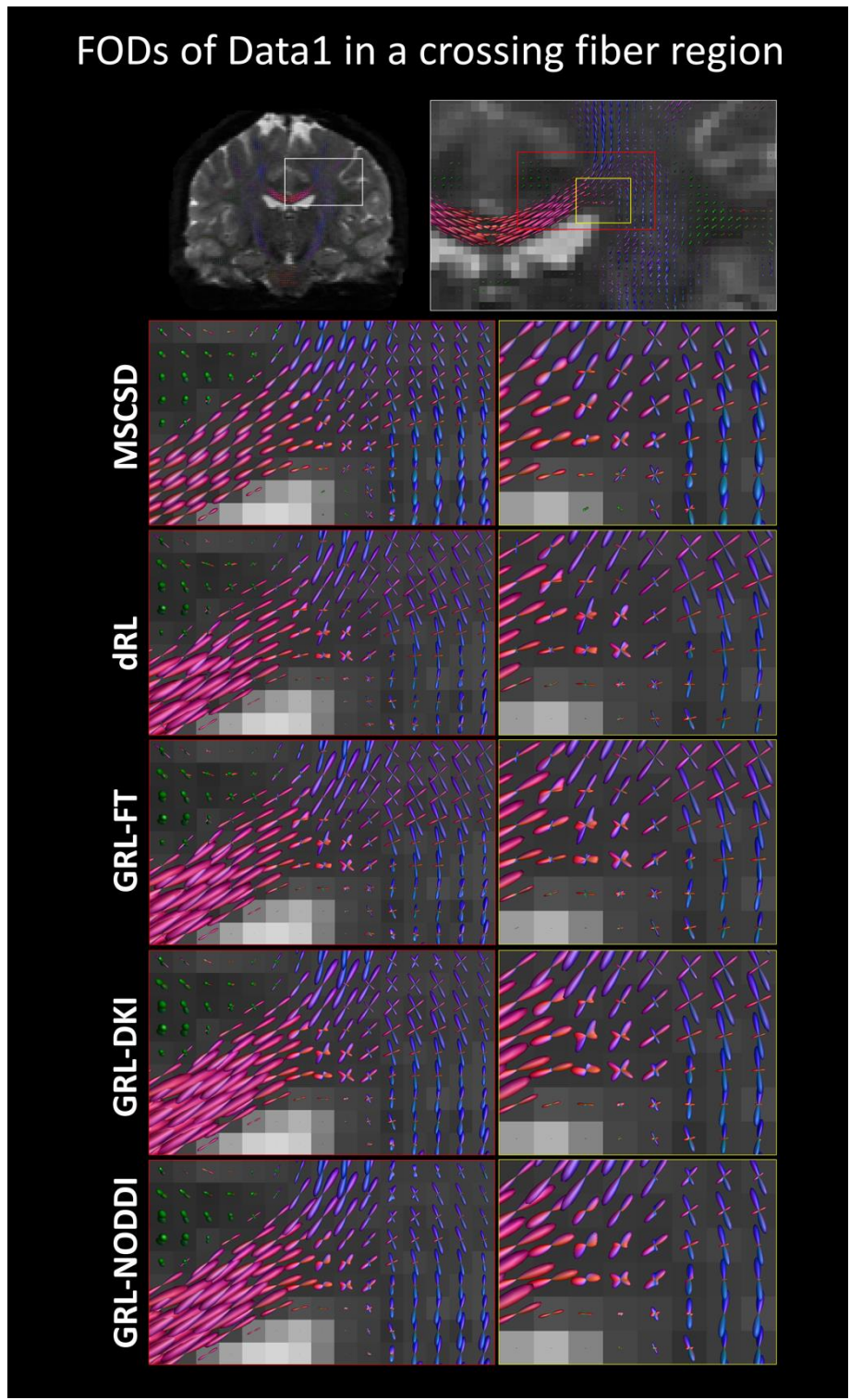

Figure 6.

An example coronal view of the FODs estimated with MSCSD, dRL and GRL on Data1 with focus on the white matter of the centrum semi-ovale. The FODs are colored encoded according to the conventional diffusion directional color scheme and overlaid on the first non-weighted image. The FODs reconstructed with dRL and GRL-FT are comparable among each other and to that derived with MSCSD, showing good separation of up to three crossing fibers (red box). In proximity of the ventricles (yellow box), GRL-DKI effectively reduces the FOD uncertainty as compared to both dRL and MSCSD. GRL-NODDI resulted in reduced resolution of crossing fibers.

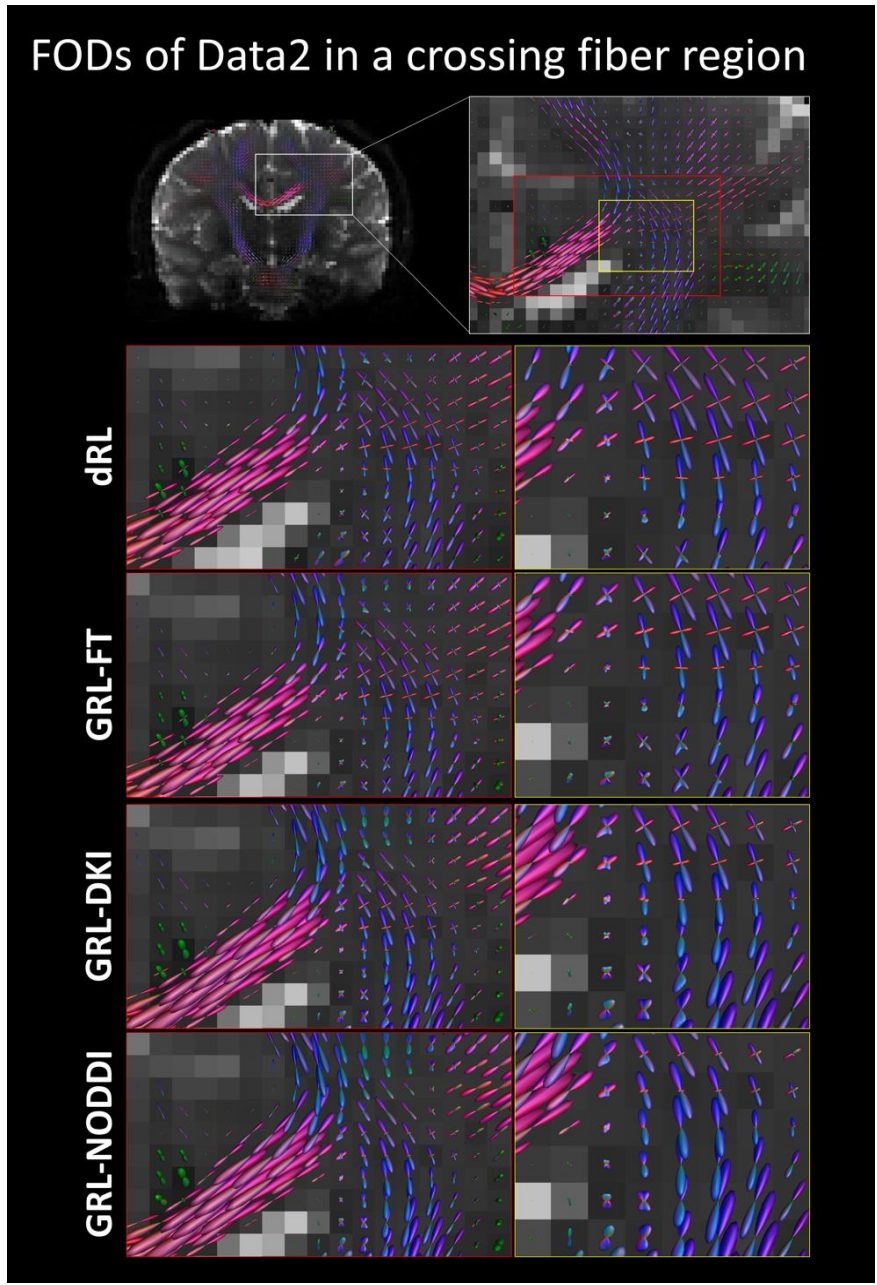

Figure 7.

An example coronal view of the FODs estimated with dRL and GRL on Data2 with focus on the white matter of the centrum semi-ovale. The FODs are colored encoded according to the conventional diffusion directional color scheme and overlaid on the first non-weighted image. In analogy with results on Data1, the FODs reconstructed with dRL, GRL-FT and GRL-DKI are comparable and show high-quality separation of up to three crossing fibers, and no spurious peaks in CSF regions.

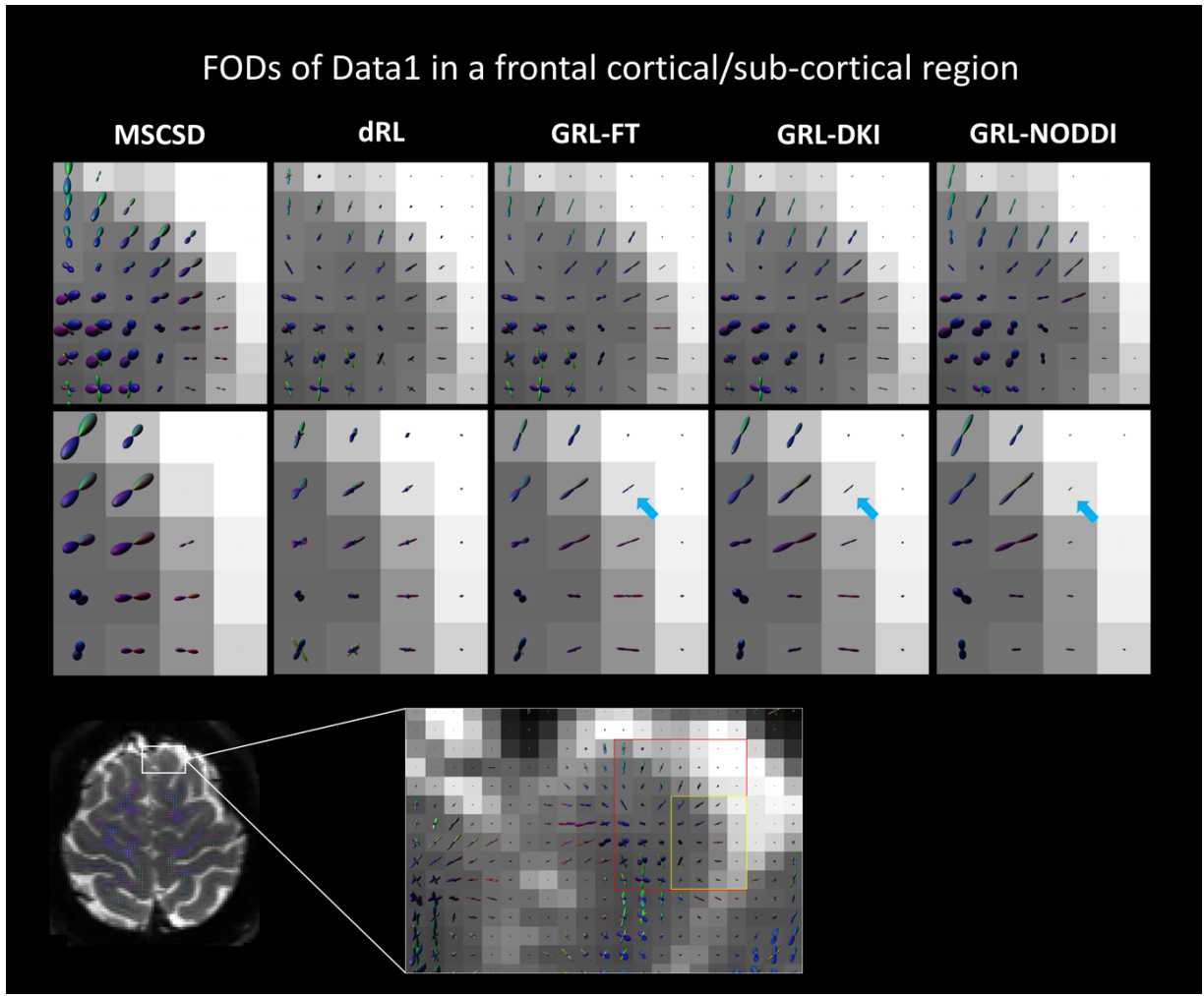

Figure 8.

An example axial view of the FODs estimated with MSCSD, dRL and GRL on Data1 with focus in a frontal region where the WM enters the GM. The FODs are colored encoded according to the conventional diffusion directional color scheme and overlaid on the first non-weighted image. GRL-FT, GRL-DKI and GRL-NODDI resulted in sharper peaks than both dRL and MSCSD, with a lower number of detected peaks in GM. The light blue arrow highlights a voxel towards the end of the cortical folding where GRL-FT and GRL-DKI estimate a sharp FOD in a radial direction with respect to the cortical folding, whereas the FOD estimated with dRL is less defined, and MSCSD provides no FOD for such voxel.

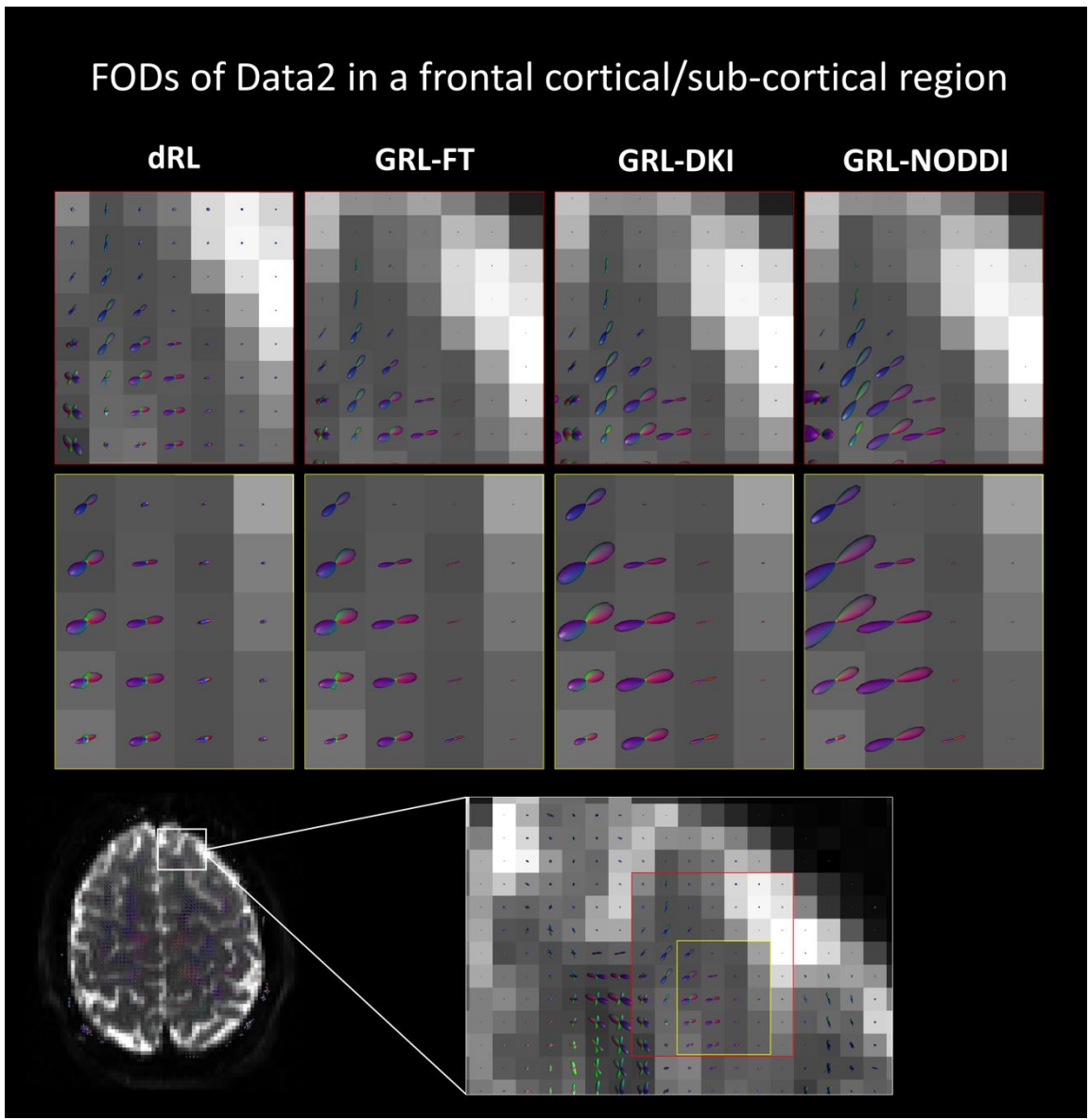

Figure 9.

An example axial view of the FODs estimated with dRL and GRL on Data2 with focus in a frontal region where the WM enters the GM. The FODs are colored encoded according to the conventional diffusion directional color scheme and overlaid on the first non-weighted image. GRL generally suppressed the effect of partial volume effects onto the FOD with higher efficacy than dRL.

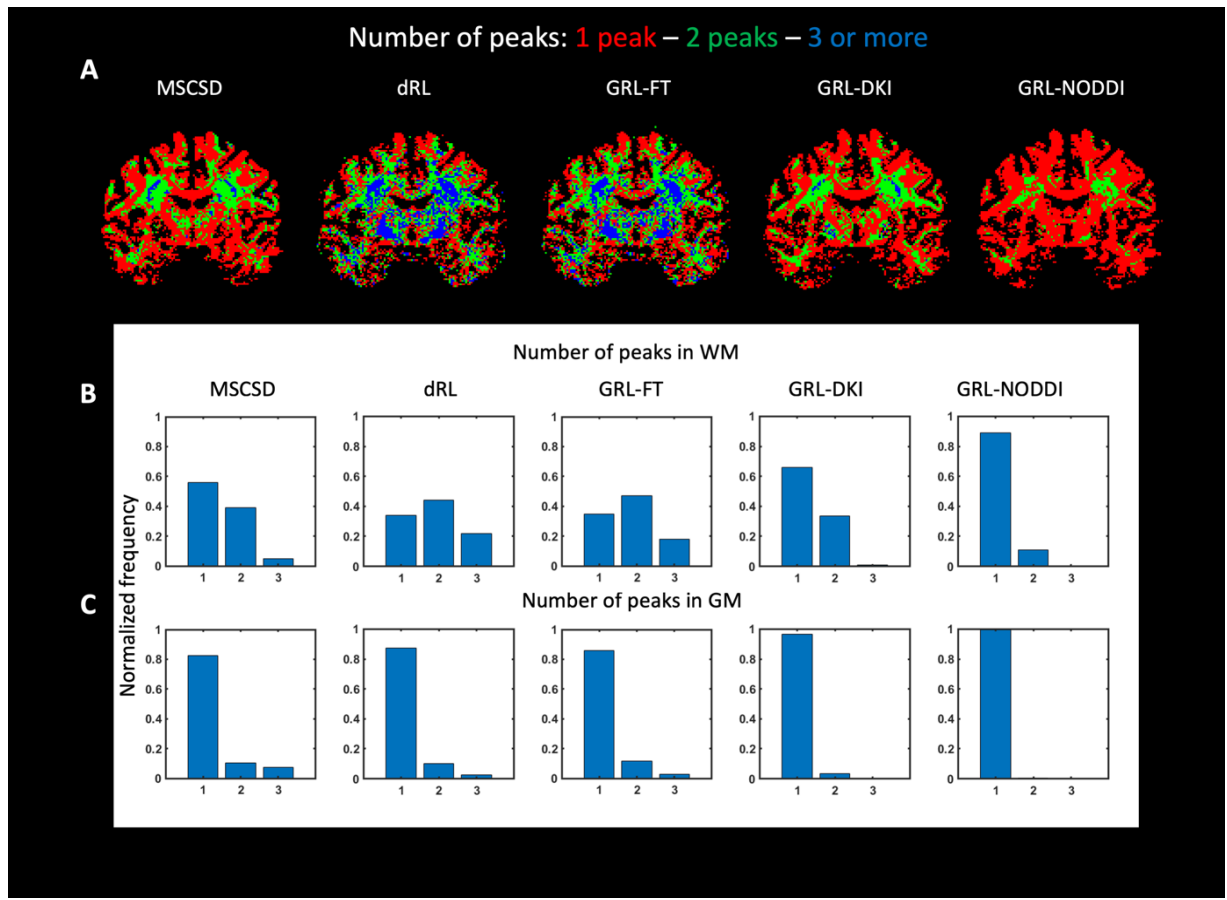

Figure 10.

The number of peaks detected with the FODs computed with MSCSD, dRL and GRL (in correspondence of different H-matrix choices) on Data1. A) an example coronal slice showing the number of detected peaks and B-C) the peak frequency in WM and GM, respectively. GRL-FT resulted in slightly reduced number of 3+ fiber crossing as compared to dRL, in particular in peri-cortical regions. The presented results are computed with a minimum peak value threshold equal to 20% the average amplitude in WM of an FOD derived with dRL.

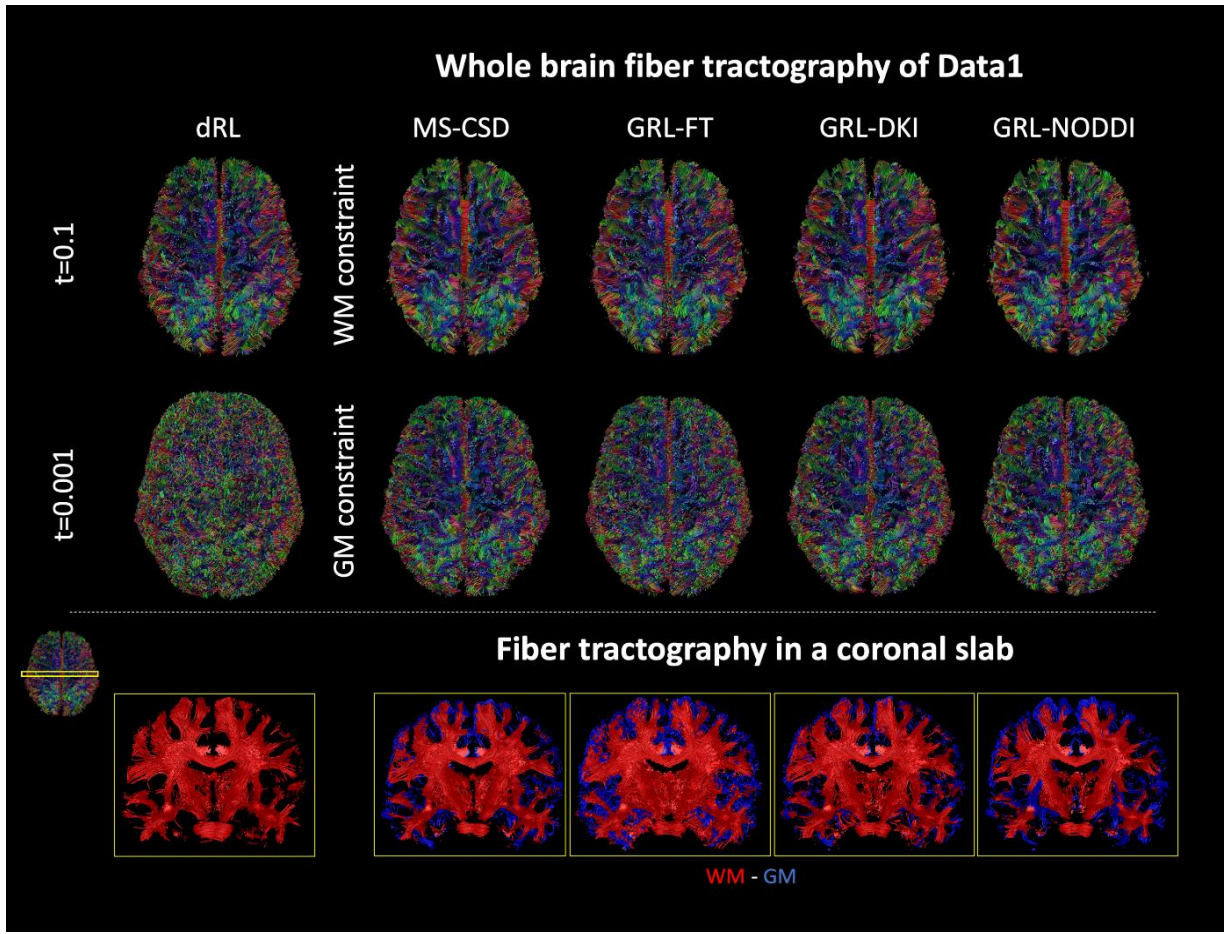

Figure 11.

Rows 1-2) the results of whole brain fiber tractography obtained with dRL, MSCSD and GRL on Data1. Fiber tractography with dRL is terminated using FOD thresholding, whereas tractography with MSCSD and GRL are free of thresholds and use the WM-like (first row) or the GM-like (second row) signal fraction maps as termination criteria. The third row shows a coronal slab of the tracking, which for GRL and MSCSD was colored according to the traversed signal fraction. Tracking with GRL-FT was comparable to dRL in WM, and tracking with GRL-DKI and MSCSD shown similar results. Conversely, with dRL it is not possible to selectively allow tracts to enter the cortex without producing a large number of spurious tracts (second row). The choice of the H-matrix has a major impact on the fiber tractography.

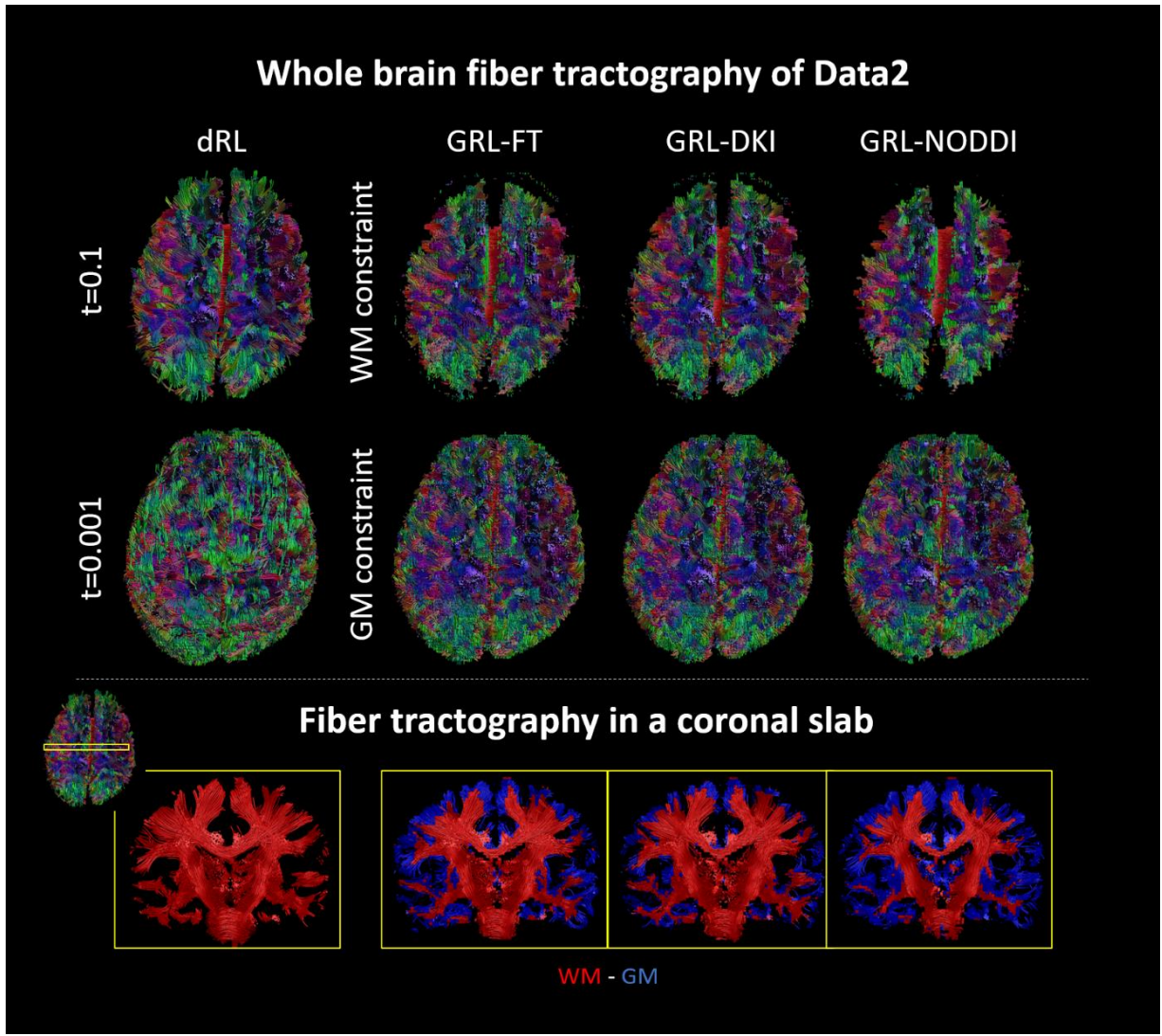

Figure 12.

Rows 1-2) the results of whole brain fiber tractography obtained with dRL and GRL on Data2. Fiber tractography with dRL is terminated using FOD thresholding, whereas tractography with GRL is free of thresholds and uses the WM-like (first row) or the GM-like (second row) signal fraction maps as termination criteria. The third row shows a coronal slab of the tracking, which for GRL was colored according to the traversed signal fraction. Tracking with GRL-FT was comparable to dRL in WM. When taken into account that the lower resolution of Data2 is likely to cause more severe partial volume effects, GRL-FT allow a finer transition into the cortical layer. Even considering tractography only in WM, the first row shows that GRL removes spurious fibers in the frontal lobe otherwise reconstructed with dRL.